\documentclass[a4paper,
               keeplastbox,   
               ]{jacow}
\usepackage{pdfpages,multirow,ragged2e} %
\makeatletter%
	\ifboolexpr{bool{xetex}}
	 {\renewcommand{\Gin@extensions}{.pdf,%
	                    .png,.jpg,.bmp,.pict,.tif,.psd,.mac,.sga,.tga,.gif,%
	                    .eps,.ps,%
	                    }}{}
\makeatother

\ifboolexpr{bool{xetex} or bool{luatex}} 
 {}                                      
 {\usepackage[utf8]{inputenc}}           

\usepackage[USenglish]{babel}

\ifboolexpr{bool{jacowbiblatex}}%
 {%
  \addbibresource{jacow-test.bib}
  \addbibresource{biblatex-examples.bib}
 }{}
\begin{document}

\title{Review of impedance-induced instabilities and their possible mitigation techniques\thanks{This project has received funding from the European Union's Horizon 2020 Research and Innovation programme under Grant Agreement No 730871.}}

\author{M. Migliorati\thanks{mauro.migliorati@uniroma1.it}, University of Rome La Sapienza and INFN-Roma1, Rome, Italy \\
		E. M\'etral, CERN, Geneva, Switzerland \\
		M. Zobov, INFN-LNF, Frascati, Italy}
	
\maketitle

\begin{abstract}
   In this paper a review of some important impedance-induced instabilities are briefly described for both the longitudinal and transverse planes. The main tools used nowadays to predict these instabilities and some considerations about possible mitigation techniques are also presented.
\end{abstract}

\section{INTRODUCTION}
The first studies of impedance-induced instabilities were developed in mid-end 60s, with the initial concepts regarding dispersion relations and coupling impedance described in the first works of V.~Vaccaro and A.~M.~Sessler~\cite{vaccaro1, vaccaro2}. In the following years, many influential researchers made the history of this important, intriguing, and always in fashion topic of particle accelerators. Over these 50 years a considerable amount of papers has been, and continues to be, published. Among them, without the intention of being exhaustive, we suggest to the reader the following references:~\cite{sacherer, chao, laclare, zotter, pellegrini, sands, laslett, courant, chin, others, emet1, emet2}.

In this paper we try to summarise the work done so far and the tools which are used nowadays to predict these impedance-induced instabilities. Of course, due to the vastness of the subject, we have to cut some of the many interesting effects that have been studied and, for others, we will just mention some aspects. Moreover, we focus only on instabilities in circular machines.

\section{SOME USEFUL DEFINITIONS}
When a beam of charged particles traverses a device which is not a perfect conductor or is not smooth, it produces electromagnetic fields that perturb the following particles. Differently from the fields generated by magnets and RF cavities, these ones depend on beam intensity and their amplitude cannot be easily changed.

\begin{figure}[!htb]
   \centering
   \includegraphics[width=.9\columnwidth]{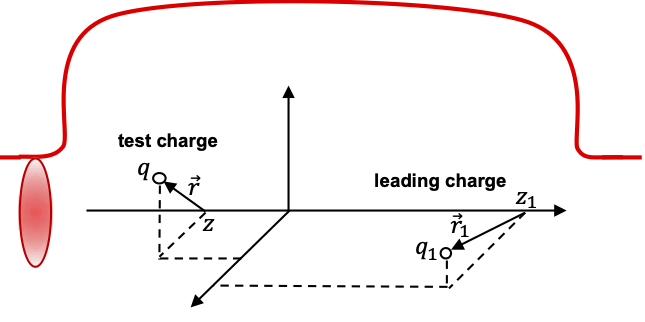}
   \caption{Sketch used for the definition of wake fields.}
   \label{figure1}
\end{figure}

These fields are generally described in time domain through the concept of wake field, or, in frequency domain, by its Fourier transform, called coupling impedance. Their importance is due to the fact that, under some conditions, they can induce instabilities. Referring to Fig.~(\ref{figure1}), let us consider two charges, a leading one (the source) $q_1$, in the position $(z_1, \vec{r}_1)$, which, interacting with an accelerator device (the red shape that we suppose of cylindrical symmetry), produces an electromagnetic field and therefore a Lorentz force on a test charge $q$ following at a distance $\Delta z=(z_1-z)$ and with a transverse displacement from the ideal orbit $\vec{r}$. 

With this geometry, by using the rigid beam approximation (the distance between the two charges remains constant inside a device) and the impulse approximation (what it cares is the impulse)~\cite{ng, palumbo}, the effects of the longitudinal and transverse components of the Lorentz force can be separated. In the longitudinal plane the effect is summarised in an energy change:
\begin{equation}
\label{eq:1}
    U(\Delta z)=\int_{device} F_{\parallel}ds \to w_{\parallel}(\Delta z)=-\frac{U(\Delta z)}{q q_1}
\end{equation}
while in the transverse plane we have a momentum kick:
\begin{equation}
\label{eq:2}
    \vec{M}(\vec{r},\Delta z)=\int_{device} \vec{F}_{\perp}ds \to \vec{w}_{\perp}(\Delta z)=\frac{1}{r}\frac{\vec{M}(\vec{r},\Delta z)}{q q_1}
\end{equation}

Here $w_{\parallel}$ and $\vec{w}_{\perp}$ are defined as the longitudinal and the transverse dipole wake functions. For Eq.~(\ref{eq:2}) we have supposed a cylindrically symmetric structure and the speed of light, otherwise also another term, called quadrupolar wake field, would have been necessary~\cite{heifets, mauro1, mauro2, mauro3}.

\section{VLASOV SOLVERS AND SIMULATION CODES}
The tools used to simulate the effects of wake fields on beam dynamics have been improved over the years, also thanks to the increased computing power. The fundamental idea to deal with these effects is quite simple: we start from the motion of a single particle inside an accelerator and include the Lorentz force due to all the others. This basic and simple idea has, however, its limits. Generally a bunch contains $10^{10} - 10^{12}$ charges, requiring the same number of equations of motion to be integrated in time. Of course, even with the computing resources available nowadays, this is still not possible. Therefore two approaches are generally used:
\begin{enumerate}
    \item at one extreme we consider a continuous distribution function describing the motion of a beam as a superposition of coherent modes of oscillation. This leads to the Vlasov (or Fokker Plank) equation (and corresponding solvers~\cite{elias,delphi});
    \item on the opposite side, we simplify the problem and reduce the number of equations by using simulation codes, which track, in time domain, about $10^6 - 10^7$ macro-particles, taking into account their electromagnetic interactions by using the concept of wake field.
\end{enumerate}
These two methods have pros and cons. For example, while Vlasov solvers in some cases may present issues related to the number of coherent modes to take into account (a convergence study is necessary), simulation codes could give non-physical results due to noise produced by the discretization with  macro-particles. Moreover, with tracking codes we can simulate any complex case while with Vlasov solvers we are limited to simpler cases. However, with tracking simulations we might miss some instabilities which would develop after the total simulated time, while with Vlasov solvers we know if the beam (some modes) will become unstable or not. It is important to remind, however, that, in parallel, simple models, as the two-particle one, have been developed to describe in a simple way some instabilities. These models allow to understand many physical aspects with quite manageable expressions.

Concerning the Vlasov equation, it describes the collective behaviour of a system of multiple particles under the influence of electromagnetic forces. Strictly speaking, the Vlasov equation is valid only for proton beams when we can ignore diffusion or damping effects. For electron, for example, synchrotron radiation cannot be neglected and, in this case, we have to use the Fokker-Plank equation that, for the longitudinal plane, can be written as~\cite{toushek}
\begin{equation}
\label{eq:3}
    \frac{\partial \psi}{\partial t}+ 
    \frac{\partial \psi}{\partial q} 
    \frac{\partial H}{\partial p} -
    \frac{\partial\psi}{\partial p}
    \frac{\partial H}{\partial q} =
    A \frac{\partial}{\partial p} (\psi p)+
    \frac{D}{2}\frac{\partial^2\psi}{\partial p^2}
\end{equation}
where $\psi$ is the longitudinal phase space distribution function, $H$ is the Hamiltonian of the system, $t$ is the time, $(q,p)$ is a set of canonical longitudinal coordinates (as, for example, time and energy offset), $A$ and $D$ depend on the synchrotron radiation and are related, respectively, to the damping and diffusion coefficients. When the left-hand side of the equation is equal to zero, i.~e.~the local particle density in phase space is constant, Eq.~(\ref{eq:3}) becomes the Vlasov equation. If we want to treat the transverse plane, we need to consider a 4D phase space and Eq.~(\ref{eq:3}) will contain other terms. 

\begin{figure*}[!htb]
\includegraphics*[width=\textwidth ]{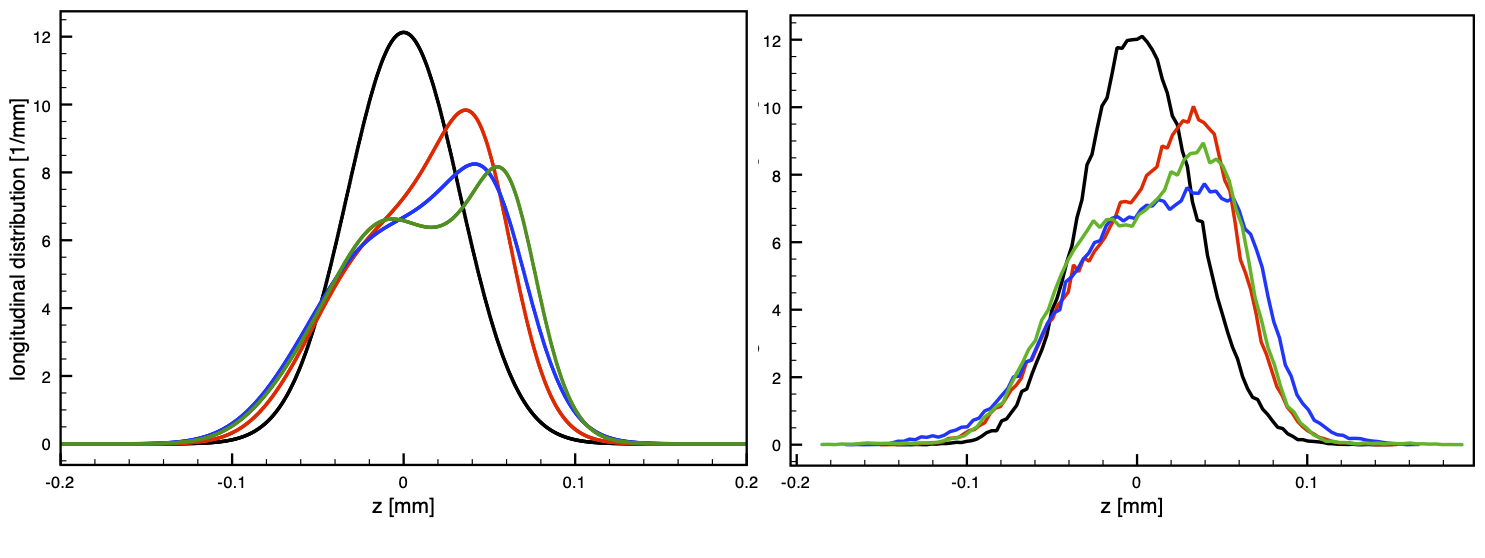}
   \caption{Longitudinal electron distribution distorted by a broad band resonator for four different intensities. Left: analytical results, right: SBSC simulation code. With increasing intensity the shape is more distorted from the unperturbed Gaussian distribution. The plot has to be considered in a qualitative way since the distortion depends not only on the intensity, but also on the impedance and other machine parameters.}
   \label{figure2}
\end{figure*}

This equation can be solved in the stationary condition: $\partial \psi / \partial t =0$. This leads to the so called Haissinski equation~\cite{haissinski}, valid for electrons, which is an integral equation that, in presence of the wake fields, gives the potential well distortion: the zero current distribution changes according to the bunch intensity and to the kind of wake field. As an example, in the left-hand side of Fig.~\ref{figure2} we have reported the solution of the equation for a broad band resonator coupling impedance at different intensities starting from an unperturbed Gaussian distribution. On the right-hand side of the same figure, the results from the simulation code SBSC~\cite{sbsc} are shown. Further examples of the bunch shape distortion due to some other impedances (resistive, capacitive and inductive) can be found in~\cite{bane}.

For proton beams, instead of the Haissinski equation, the solution of the Vlasov stationary equation $\psi_0(q,p)$ is any function of the Hamiltonian $H_0(q,p)$. On its turn, $H_0(q,p)$ depends on the wake fields and, in the end, we have, similarly to electrons, a potential well distortion.

Instead of a continuous function, simulation codes track, in time domain, some macro-particles representing the whole bunch. Let us consider, for simplicity, only the longitudinal plane, with an energy exchange due to a single RF system, the wake field and the synchrotron radiation. Similar equations can also be written in the transverse plane.
Under these conditions, for each macro-particle $i$, we have two equations, which can be written as
\begin{multline}
    \Delta \varepsilon_i =\frac{q V_{RF} (\sin \varphi_i-\sin \varphi_s)}{E_s} +\\ 
    \frac{ -qV_{wf}(\varphi_i)+R(T_0)}{E_s} - 2\frac{T_0}{\tau_s}\varepsilon_i
\end{multline}
\begin{equation}
    \Delta(\varphi_i-\varphi_s)=-\frac{2\pi h \eta}{\beta^2}\varepsilon_i
\end{equation}
where $\Delta$ means the variation of a given quantity in one integration step $T_0$ (that can be one revolution turn for example), $\varepsilon_i$ is the normalised energy difference with respect to the synchronous particle, $q$ is the particle charge, $V_{RF}$ is the RF peak voltage, $\varphi_i$ is the particle phase with respect to the RF, $\varphi_s$ is the synchronous phase, $R$ is a stochastic variable changing at each integration step and taking into account the fact that the electromagnetic radiation occurs in quanta of discrete energy, $E_s$ is the synchronous particle energy, $\tau_s$ is the longitudinal damping time, $h$ is the harmonic number, $\eta$ is the slippage factor, $\beta$ is the relativistic velocity factor, and the effect of the wake field, which couples the equations to those of all the other macro-particles, is given by the wake induced voltage
\begin{equation}
\label{eq:6}
    V_{wf}(\varphi_i)=\frac{Q_{tot}}{N_m}\sum_{j=1}^{N_m}w_{\parallel} (\varphi_i-\varphi_j)
\end{equation}
with $Q_{tot}$ the total bunch charge and $N_m$ the number of macro-particles. We have also considered that $\varphi_i-\varphi_s>0$ for a particle behind the synchronous one, that is with positive time delay. 

Since Eq.~(\ref{eq:6}) has to be evaluated for each of the $N_m$ macro-particles, then $(N_m-1)N_m/2$ operations are needed for each time step. In order to reduce the computing time and only for the evaluation of the wake field effects, the bunch is generally divided into $N_s$ slices $(N_s < N_m)$, and the wake induced voltage is calculated at the centre of each slice $i'$ such that
\begin{equation}
\label{eq:7}
    V_{wf}(\varphi_{i'})=\frac{Q_{tot}}{N_m}\sum_{j'=1}^{N_s}n_{j'} w_{\parallel} (\varphi_{i'-}\varphi_{j'})
\end{equation}
with $n_{j'}$ the number of macro-particles in the slice $j'$. To obtain the induced voltage for each macro-particle, an interpolation on the Eq.~\ref{eq:7} is used. As shown in the right-hand side of Fig.~\ref{figure2}, simulation codes, for the stationary case, give the same results as the analytical approach.

\section{INSTABILITIES IN CIRCULAR ACCELERATORS}

For the study of instabilities in circular accelerators, it is convenient to separate the longitudinal and transverse planes as we did for the wake field described by Eqs.~(\ref{eq:1}) and (\ref{eq:2}). Moreover, for each plane, we generally distinguish between the single-bunch effects, generated by the short-range wake field, which has, in the corresponding frequency domain, coupling impedances with a poor frequency resolution (broad band impedance), and the multi-bunch (or multi-turn) effects produced by long-range wake fields or, in frequency domain, by high quality (often unwanted) resonant modes.

\begin{figure*}[!htb]
   \centering
   \includegraphics*[width=.99\textwidth]{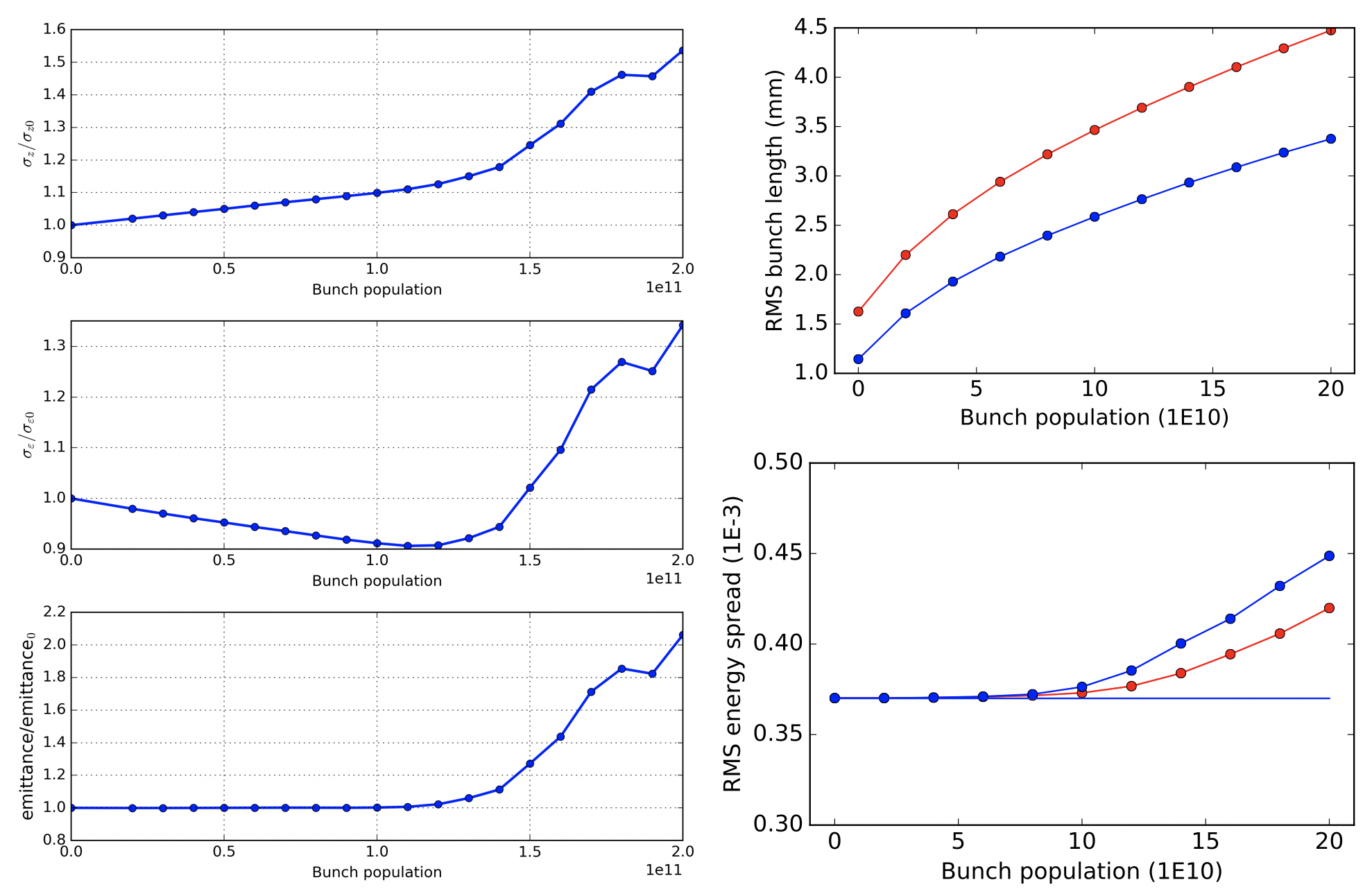}
   \caption{Bunch length, energy spread and longitudinal emittance vs bunch intensity for a case with protons (left). Bunch length and energy spread vs bunch intensity for electrons (right) for two different initial bunch lengths.}
   \label{figure3}
\end{figure*}

Let us first consider single-bunch effects at low intensity in the longitudinal plane. As already discussed in the previous section, the effect in this case is a distortion of the distribution function that depends on the bunch current. There exists a bunch distribution which corresponds to the stationary solution of the Vlasov, or the Fokker-Plank, equation. There is a different behaviour between protons and electrons since, in the first case, we can neglect the effects of synchrotron radiation, and the bunch length and energy spread change with intensity in such a way to preserve the longitudinal emittance, as shown in Fig.~\ref{figure3}, left-hand side. For electrons, instead, shown in the right-hand side of the same figure for two different initial bunch lengths, the energy spread remains constant, due to an equilibrium between radiation damping and quantum fluctuations noise, while the potential well distortion changes the bunch length (and shape).

In both cases we can observe an intensity threshold above which the longitudinal emittance (for protons) or the energy spread (for electrons) start to increase. Above this threshold we are in the so called microwave instability regime, characterised by an anomalous increase of bunch length and energy spread. In some cases, longitudinal oscillations of the bunch are observed (no stationary solution exists). However, for this kind of longitudinal instability, typically there are no beam losses. 

To study analytically the single-bunch instabilities in the longitudinal plane (but the same method is also valid in the transverse plane), the steps to do can be summarised as follows:
\begin{enumerate}
    \item use a perturbation method and write the phase space distribution as $\psi(q,p;t)=\psi_0(q,p)+\Delta \psi(q,p;t)$;
    \item use, as canonical longitudinal coordinates, the action-angle coordinates $(I,\phi)$ and consider the perturbation as sum of azimuthal coherent modes $R_m(I)$ oscillating with an unknown coherent frequency $\Omega$:
    \begin{equation}
        \Delta \psi(q,p;t) =\sum_{m=-\infty}^{\infty}R_m(I)e^{im\phi} e^{-i \Omega t};
    \end{equation}
    \item consider the instability produced by the wake fields excited only by the perturbation (not by the stationary distribution);
    \item from the Valsov equation, the so-called Sacherer integral equation is then obtained (the multi-bunch case can be treated in a similar way);
    \item solve the integral equation: there are several methods to obtain the solution~\cite{laclare}. For example it is possible to expand each azimuthal mode $R_m(I)$ in terms of a set of orthonormal functions $g_{mk}(I)$ with unknown amplitude $\alpha_{mk}$ and a proper weight function $W(I)$ which depends on (the derivative of) the stationary distribution:
    \begin{equation}
    \label{eq:9}
        R_m(I)=W(I)\sum_{k=0}^{\infty}\alpha_{mk}g_{mk}(I);
    \end{equation}
    \item from Eq.~(\ref{eq:9}) an infinite set of linear equations is obtained. The eigenvalues represent the coherent frequencies and the eigenvectors the corresponding modes:
    \begin{equation}
        (\Omega -m \omega_s)\alpha_{mk}=\sum_{m'=-\infty}^{\infty} \sum_{k'=0}^{\infty}M_{kk'}^{mm'}\alpha_{m'k'}.
    \end{equation}
\end{enumerate}

\begin{figure}[!htb]
   \centering
   \includegraphics[width=\columnwidth]{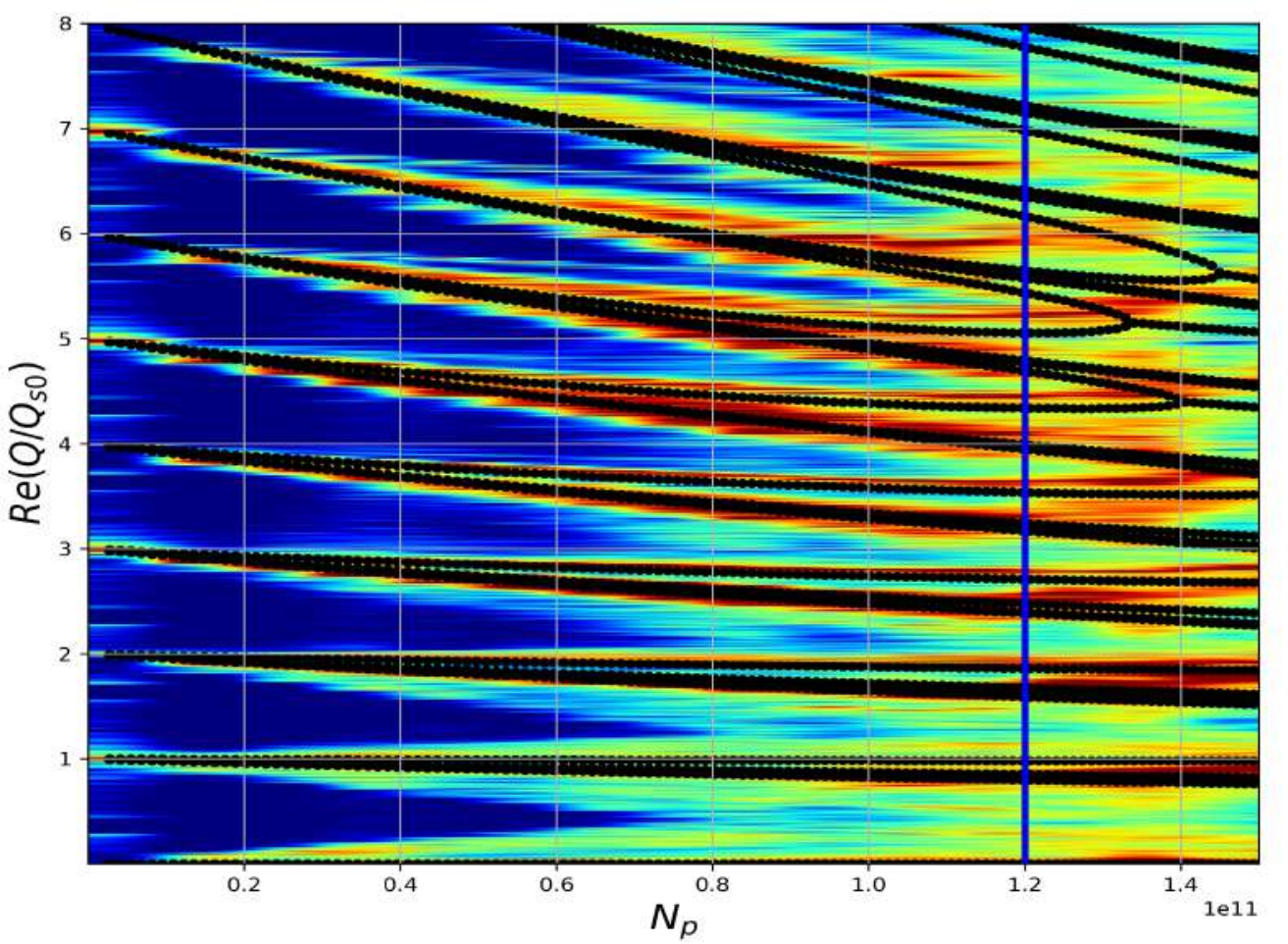}
   \caption{Real part of longitudinal coherent frequency normalised to the synchrotron frequency vs bunch intensity for a broad band resonator impedance.}
   \label{figure4}
\end{figure}

For low intensity, we ignore the coupling of radial modes that belong to different azimuthal families $(m=m')$, the matrix of the eigenvalue system is Hermitian, the eigenvalues are always real and no instability occurs (this is true only in longitudinal plane). Only coupled-bunch instabilities (interaction with high Q resonators) can occur if we consider single azimuthal modes. At high intensity, however, mode coupling can occur by taking into account different azimuthal modes. An example of this behaviour is shown in Fig.~\ref{figure4}, where we have reported, in black, the coherent frequencies of the first azimuthal modes as a function of bunch intensity, as given by GALACLIC Vlasov solver~\cite{elias2} for a broad band resonator impedance. From the figure, with the parameters used for this case, we observe a mode coupling of modes 6 and 7 around 1.3$\times 10^{11}$ particles per bunch.

In the same figure we have also shown the results from a simulation code~\cite{music} which predicts a similar behaviour even with some small differences. It is important to stress, however, that, for the Vlasov solver, we considered here the simplest model of potential well distortion where the synchronous phase shift vs bunch intensity is neglected, i.e. the shape of the longitudinal distribution is conserved. The effect of the full potential well distortion should be studied in the future.

For proton machines, the synchrotron tunes are in general much smaller than those in electron machines. As a consequence, when considering collective instabilities, in some cases the synchrotron period of protons can be neglected because it is much longer than the instability growth times. Moreover, the wavelength of the perturbation producing the instability is often of the size of the radius of the vacuum chamber, which is usually much shorter than the length of the proton bunch. Therefore, proton bunches, in some cases, can be viewed locally as coasting beams in many instabilities considerations. Boussard~\cite{boussard} suggested to apply the same criterion of coasting beams (Keil-Schnell)~\cite{keil} to bunched beams, obtaining a threshold current of
\begin{equation}
    I_{th}=\frac{\sqrt{2\pi}|\eta|(E_s/e)\sigma_{\varepsilon}^2 \sigma_z}{R |Z_{\parallel}/n|}
\end{equation}
where $\sigma_{\varepsilon}$ is the RMS energy spread, $\sigma_z$ the RMS bunch length, and $|Z_{\parallel}/n|$ the coupling impedance evaluated at the $n^{th}$ harmonic of the revolution frequency. In this case the microwave instability is not due to a mode coupling but each single revolution harmonic can be considered as an independent mode. The Boussard criterion can be a good indicator on how to cope with the microwave instability and where to act to mitigate such effect.

For the transverse plane the procedure is similar to the longitudinal one with few differences:
\begin{itemize}
    \item the bunch is supposed to have only a dipole moment in the transverse plane;
    \item this dipole moment is not constant longitudinally, but it has a structure which depends on the longitudinal mode number $m$;
    \item the modes are called transverse modes, but the transverse structure is a pure dipole and the main task is to find their longitudinal structure;
    \item the Vlasov equation needs to take into account both the transverse and the longitudinal phase spaces. Fortunately, however, in several cases, the transverse structure of the beam is simple.
\end{itemize}

\begin{figure}[!htb]
   \centering
   \includegraphics[width=\columnwidth]{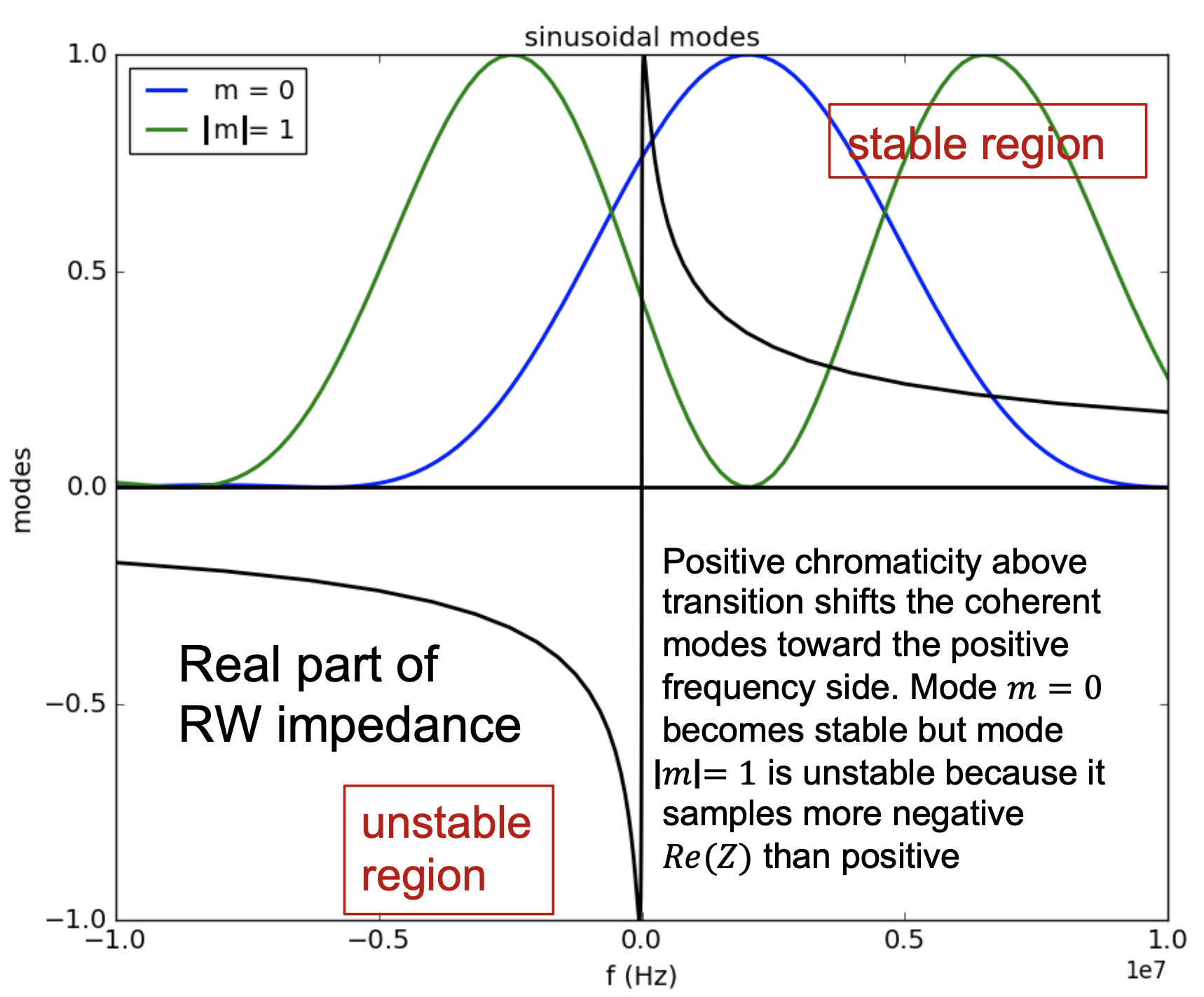}
   \caption{Sketch of the real part of a resistive wall impedance vs frequency together with the first two coherent modes of oscillation.}
   \label{figure5}
\end{figure}

The eigenvalue system that is obtained from the Vlasov equation in the transverse plane is
\begin{equation}
    (\Omega -\omega_{\beta}-m \omega_s)\alpha_{mk}=\sum_{m'=-\infty}^{\infty} \sum_{k'=0}^{\infty}M_{kk'}^{mm'}\alpha_{m'k'},
\end{equation}
with $\omega_{\beta}$ the angular betatron frequency. The matrix elements, in this case, depend also on chromaticity. When chromaticity is zero, similarly to the longitudinal plane, the only instability for low intensity beams is due to high Q resonators. However, if the chromaticity is different from zero, differently from the longitudinal plane, single azimuthal modes can be unstable producing the so called head-tail instability. This is not an intensity threshold mechanism, and it is due to the coupling of the real part of the transverse impedance at negative frequency with the coherent modes shifted from the origin due to the chromaticity, as shown, for example, in Fig.~\ref{figure5} for the resistive wall impedance. In the figure, indeed, a positive chromaticity above transition shifts the coherent modes toward the positive frequency side. In this situation, the mode $m=0$ becomes stable but mode $|m|=1$ is unstable because it samples a real part of the impedance in the negative frequency range higher than that at positive frequencies.

\begin{figure}[!htb]
   \centering
   \includegraphics[width=1\columnwidth]{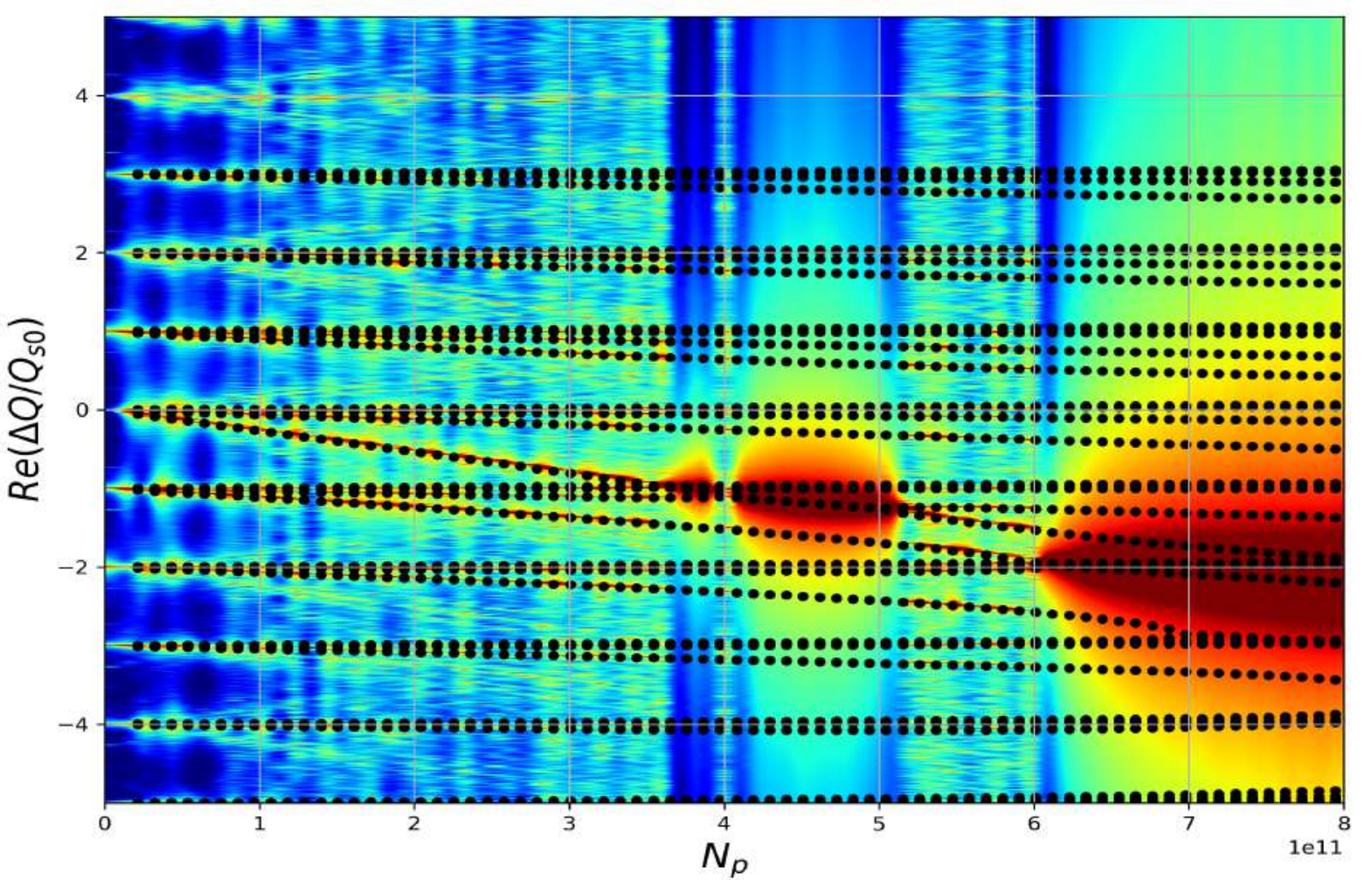}
   \caption{Real part of transverse coherent frequency normalised to the synchrotron frequency vs bunch intensity.}
   \label{figure6}
\end{figure}

 In addition to the head tail instability, at high intensity, mode coupling can occur for zero chromaticity, as shown in Fig.~\ref{figure6} where we have reported, as for the longitudinal case, a comparison between the GALACTIC Vlasov solver~\cite{elias2} and the PyHEADTAIL simulation code~\cite{PyHT}. In this case an excellent agreement has been reached (for both the real and imaginary parts of the coherent frequency shifts).

\section{some considerations about mitigation techniques}

A very effective way to mitigate any kind of impedance-induced instability is, of course, that of reducing the machine coupling impedance. This can be achieved, for example, by tapering abrupt transitions, by avoiding electrical discontinuities, by shielding unwanted parasitic cavities, and so on. However, there could be cases in which these measures are not possible (or are insufficient). As further comments about mitigation, in the longitudinal plane we observe that no feedback systems can be used to suppress the microwave instability. However, the Boussard criterion can give important indications on how to cope with this instability. For example the increase of momentum compaction (which can be considered a strong factor), of energy spread (heating the bunch, e.g. with wigglers in electron machines) are effective means to increase the instability threshold. It is also important to note that some machines work in the microwave instability regime. For the transverse plane, in general the lattice choice is quite a strong factor to mitigate the instabilities by acting on: tunes, linear and nonlinear chromaticity, coupling, tune dependence on the oscillation amplitude etc. Feedback systems can be used for both proton and electron beams, and they are working very well for coupled-bunch instabilities, as discussed below.

\begin{figure*}[!htb]
   \centering
   \includegraphics*[width=\textwidth]{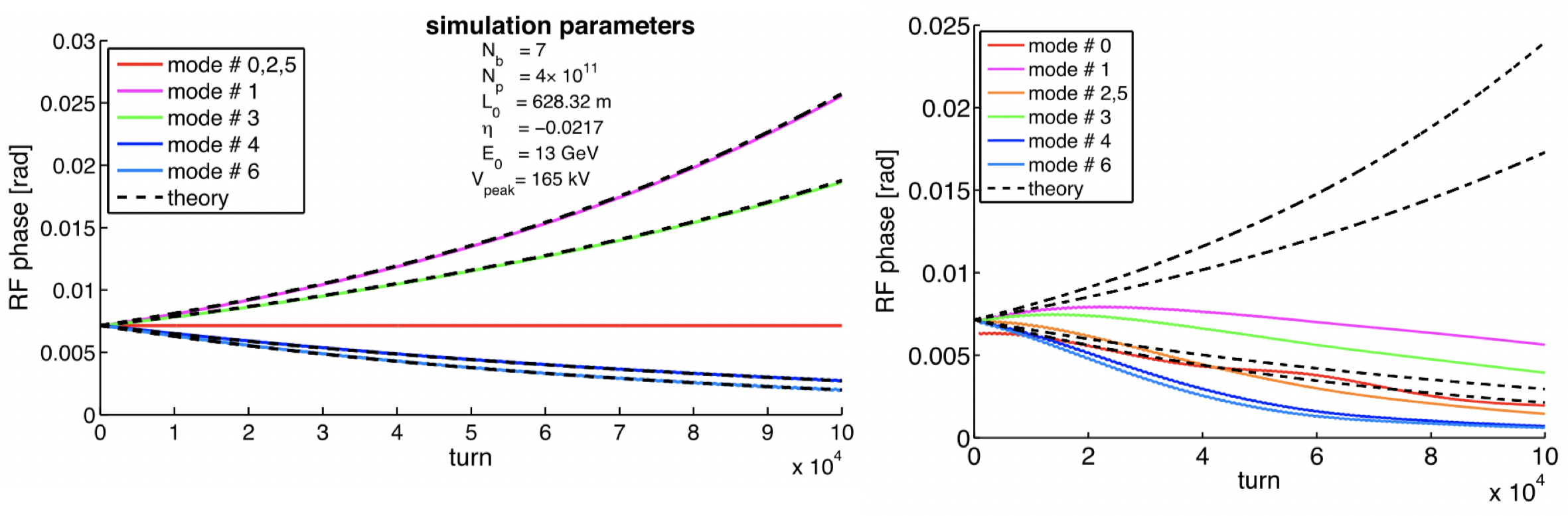}
   \caption{Growth rates of coupled-bunch modes as a function of time (number of turns) for a multi-bunch instability. In the left figure the bunch is supposed a point charge, on the right-hand side a finite length is given.}
   \label{figure7}
\end{figure*}

A particular mention needs the coupled-bunch instability due to high quality resonant modes. The analytical treatment is similar to the single-bunch case, with an additional index in the coherent modes taking into account the coupled-bunch oscillations. This kind of instability leads to a loss of the beam in both planes if mitigation techniques are not used. For example in Fig.~\ref{figure7} we show the growth rates of the coherent coupled-bunch modes for a case with 7 bunches (modes from 0 to 6) by considering a bunch as a point charge (left-hand side). 

We can see that some coherent modes are unstable, others are not excited, and others are stable. If we use a bunch with a given length (right-hand side), due to a spread in the synchrotron tunes within a bunch caused by the non linearities of RF system, there is a Landau damping stabilizing the modes~\cite{music}. The effect, of course, depends on bunch length. Other mitigation techniques for this instability consist in damping unwanted resonant high quality modes, in using a feedback system, in recurring to a higher harmonic cavity (Landau cavity), both active or passive, in recurring to a RF voltage modulation which creates non-linear resonances which redistribute the longitudinal distribution in phase space reducing the density in the bunch core and thus decoupling the multi-bunch motion, or, finally, in recurring to uneven fill of the beam~\cite{mbunch} which changes the bunch spectrum. 

Finally, we observe that sometimes interplay with other effects can have a beneficial role in suppressing the impedance related instabilities. For example, the Landau damping due to beam-beam interaction helps in the damping of both transverse~\cite{minty, buffat} and longitudinal~\cite{drago} instabilities, and in the future supercolliders, FCC-ee~\cite{fccee} and CEPC~\cite{cepc}, the energy spread due to beamstrahlung in beam-beam collision helps increasing the microwave instability threshold~\cite{fccrw}.

More details about mitigation techniques can be found in the talks of this Workshop.

\section{CONCLUSIONS}
In this paper some impedance-induced instabilities have been shortly reviewed. We focused principally on the longitudinal and transverse single-bunch instabilities. However the subject is very broad and this short discussion cannot do justice of the high quality and large amount of work that has been done since the first pioneering works of mid-end 60s. A short, non exhaustive list of arguments that have not been touched is the following: coasting beam instabilities (as negative mass instability), not relativistic beams, space charge effects (which are not strictly impedance-induced instabilities), Landau damping and dispersion integrals, saw-tooth instabilities for electrons, Robinson’s instability, transition crossing, impedance effects in LINACS, such as the beam break-up instability, the microbunching instability in RF and magnetic compressors, other impedance-induced effects which are not real instabilities but can influence the machine performances (as the effect of detuning impedance, beam energy spread in LINACS and so on).

The subject of impedance-induced instability is one of the main topics for modern high performance accelerators. Even if the roots of this subject are more than 50 years old, it is still a cutting-edge in the beam physics. Many researchers have been working over the years on this subject and very elegant and well-established theories have been proposed explaining many experimental observations. In some cases we still need to study in more detail the interplay among different mechanisms (e.g. with optics) and in particular we need to better understand some mitigation techniques.

The best proof about our comprehension of these instabilities is that particle accelerators work and are successful. After 50 years, this couldn’t be only a coincidence. However, there are still “dark sides” that have to be illuminated by the young generation, which, we hope, will continue the work with the passion that has marked so far the protagonists of this fascinating subject.

\section{ACKNOWLEDGEMENTS}
 This work was partially supported by the European Commission under the HORIZON 2020 Integrating Activity project ARIES, Grant agreement No. 730871, and by INFN National committee V through the ARYA project.

%
%
\ifboolexpr{bool{jacowbiblatex}}%
	{\printbibliography}%

\begin{thebibliography}{9} 
	
	\bibitem{vaccaro1}
		V. G. Vaccaro, CERN Technical Report No. ISR-RF/66-35, 1966.

	\bibitem{vaccaro2}
		A. M. Sessler and V. G. Vaccaro, CERN Technical Report No. CERN 67-2, 1967.

	\bibitem{sacherer}
	F. Sacherer, IEEE Trans. Nucl. Sci. 20, 825, 1973; 24,
    1393, 1977.

    \bibitem{chao}
    A. W. Chao, Physics of Collective Beam Instabilities in
    High Energy Accelerators, John Wiley \& Sons, New York,
    1993.
    
    \bibitem{laclare}
    J.L. Laclare, CERN Report No. 87-03, Vol. I, p. 264,
    Geneva, Switzerland, 1987.
    
    \bibitem{zotter}
    B. Zotter, CERN Reports No. CERN-SPS/81-18 (DI), No. CERN-SPS/81-19 (DI), and No. CERN-SPS/81-20 (DI), Geneva, Switzerland, 1981.
    
    \bibitem{pellegrini}
    C. Pellegrini, AIP Conf. Proc. 87, 77 (1982).
    
    \bibitem{sands}
    M. Sands, Reports No. SLAC-TN-69-8 and No. SLAC-TN-69-10, 1969.
    
    \bibitem{laslett}
    L. J. Laslett, V. K. Neil, and A. M. Sessler, Rev. Sci. Instrum., vol. 36, no. 4, pp. 436–448, 1965.
    
    \bibitem{courant}
    E. D. Courant and A. M. Sessler, Rev. Sci. Instrum., vol. 37, no. 11, pp. 1579–1588, 1966.
    
    \bibitem{chin}
    Y. H. Chin and K. Satoh, Nucl. Instrum. Methods Phys. Res. A, no. 207, pp. 309–320, 1983.
    
    \bibitem{others}
    See, e.g., T. Suzuki and K. Yokoya, Nucl. Instrum. Methods Phys. Res. 203, 45, 1982.
    
    \bibitem{emet1}
    E. M\'etral et al., IEEE Transactions on Nuclear Science, Vol. 63, No. 2, 50 p, April 2016.
    
    \bibitem{emet2}
    E. M\'etral (Issue Editor), ICFA Beam Dynamics Newsletter No. 69 devoted to the Collective Effects in Particle Accelerators, 310 p, December 2016.
    
    \bibitem{ng}
    K. Y. Ng, Physics of Intensity Dependent Beam Instabilities, World Scientific, Singapore, 2006.
    
    \bibitem{palumbo}
    L. Palumbo, V. G. Vaccaro, and M. Zobov, CERN Technical Report No. CERN 95-06, 1995 [arXiv:physics/0309023].
    
    \bibitem{heifets}
    S. Heifets, A. Wagner, B. Zotter, SLAC/APll0, January 1998.
    
    \bibitem{mauro1}
    S. Persichelli, N. Biancacci, M. Migliorati, L. Palumbo, and V.G. Vaccaro, Phys. Rev. Accel. Beams 20, 101004, 2017.

    \bibitem{mauro2}
    M. Migliorati, N. Biancacci, M. R. Masullo, L. Palumbo, and V. G. Vaccaro, Phys. Rev. Accel. Beams 21, 124201, 2018.

    \bibitem{mauro3}
    M. Migliorati, L. Palumbo, C. Zannini, N. Biancacci, and V. G. Vaccaro, Phys. Rev. Accel. Beams 22, 121001, 2019.
    
    \bibitem{elias}
    E. M\'etral, Proc. 10th Int. Particle Accelerator Conf. IPAC2019, Melbourne, Australia, 19-24 May 2019, pp. 312-315. doi:10.18429/JACoW-IPAC2019-MOPGW087
    
    \bibitem{delphi}
    N. Mounet, DELPHI: an analytic Vlasov solver for impedance-driven modes (2014), CERN-ACC-SLIDES-2014-0066 \url{http://cds.cern.ch/record/ 1954277/files/CERN-ACC-SLIDES-2014-0066.pdf}
    
    \bibitem{toushek}
    C. Bernardini, and B. Touschek: On the quantum losses in an electron synchrotron, Laboratori Nazionali di Frascati de1 CNEN, Nota interna No. 34 (April 1960).
    
    \bibitem{haissinski}
    J. Ha\"issinski, Il Nuovo Cimento B 18, 72, 1973.
    
    \bibitem{sbsc}
    M. Migliorati et al., Phys. Rev. ST Accel. Beams, vol. 16, p. 031001, March 2013.
    
    \bibitem{bane}
    K. L. F. Bane, SLAC-PUB-5177 February 1990 (A).
    
    \bibitem{elias2}
    E. M\'etral, M. Migliorati, "Vlasov solvers and simulation code analysis for mode-coupling instabilitites in both longitudinal and transverse planes", these proceedings.
    
    \bibitem{music}
    M. Migliorati, L. Palumbo, Phys. Rev. ST Accel. Beams, vol. 18, p. 031001, March 2015. doi:10.1103/PhysRevSTAB.18.031001
    
    \bibitem{boussard}
    D. Boussard, CERN LabII/RF/75-2, 1975.
    
    \bibitem{keil}
    E. Keil, W. Schnell, CERN Report ISR-TH-RF/69-48, 1969.
    
    \bibitem{PyHT}
    K. Li, “Headtail code,” 2015. \url{https://indico.cern.ch/event/378615/}, [Online; accessed 16.05.2019].
    
    \bibitem{mbunch}
    S. Prabhakar, J. D. Fox, and D. Teytelmann, Phys. Rev. Lett., vol. 86, pp. 2022-2025, Mar. 2001.
    
    \bibitem{minty}
    M.G. Minty et al., Rep. SLAC-PIB-8363, Feb. 2000. https://www.slac.stanford.edu/cgi-bin/getdoc/slac-pub-8363.pdf
    
    \bibitem{buffat}
    X. Buffat et al., Stability diagrams of colliding beams in the Large Hadron Collider, PRSTAB 17 111002 2014.
    
    \bibitem{drago}
    A. Drago, P. Raimondi, M. Zobov, D. Shatilov, Phys. Rev. ST Accel. Beams vol. 14, p. 092803, 2011. doi:10.1103/PhysRevSTAB.14.092803
    
    \bibitem{fccee}
    The FCC Collaboration, FCC-ee: The Lepton Collider: Future Circular Collider Conceptual Design Report Volume 2, European Physical Journal Special Topics, 228, 261-623, 2019, doi: 10.1140/epjst/e2019-900045-4.
    
    \bibitem{cepc}
    CEPC Study Group, CEPC Conceptual Design Report: Volume 1 - Accelerator, 2018. 	arXiv:1809.00285
    
    \bibitem{fccrw}
    M. Migliorati, E. Belli, M. Zobov, Phys. Rev. Acc. and Beams, vol. 21, p. 041001, 2018. \\
    doi: 10.1103/PhysRevAccelBeams.21.041001.


	\end{thebibliography}
	{%
	
	
} 
%
%


\end{document}